# Incorporating human and learned domain knowledge into training deep neural networks: A differentiable dose volume histogram and adversarial inspired framework for generating Pareto optimal dose distributions in radiation therapy

Dan Nguyen, Rafe McBeth, Azar Sadeghnejad Barkousaraie, Gyanendra Bohara, Chenyang Shen, Xun Jia, and Steve Jiang

Medical Artificial Intelligence and Automation (MAIA) Laboratory,
Department of Radiation Oncology, UT Southwestern Medical Center, Dallas, TX, USA
Dan.Nguyen@UTSouthwestern.edu

**Purpose**: We propose a novel domain specific loss, which is a differentiable loss function based on the dose volume histogram, and combine it with an adversarial loss for the training of deep neural networks. In this study, we trained a neural network for generating Pareto optimal dose distributions, and evaluate the effects of the domain specific loss on the model performance.

**Methods**: In this study, 3 loss functions—mean squared error (MSE) loss, dose volume histogram (DVH) loss, and adversarial (ADV) loss—were used to train and compare 4 instances of the neural network model: 1) MSE, 2) MSE+ADV, 3) MSE+DVH, and 4) MSE+DVH+ADV. The data for 70 prostate patients, including the planning target volume (PTV), and the organs-at-risk (OAR) were acquired as 96 x 96 x 24 dimension arrays at 5 mm$^3$ voxel size. The dose influence arrays were calculated for all 70 prostate patients using a 7 equidistant coplanar beam setup. Using a scalarized multicriteria optimization for intensity modulated radiation therapy, 1200 Pareto surface plans per patient were generated by pseudo-randomizing the PTV and OAR tradeoff weights. With 70 patients, 84,000 plans were generated in total. We divided the data into 54 training, 6 validation, and 10 testing patients. Each model was trained for a total of 100,000 iterations with a batch size of 2. All models used the Adam optimizer with a learning rate of $1 \times 10^{-3}$.

**Results**: Training for 100,000 iterations took 1.5 days (MSE), 3.5 days (MSE+ADV), 2.3 days (MSE+DVH), 3.8 days (MSE+DVH+ADV). After training, the prediction time of each model is 0.052 seconds. Quantitatively, the MSE+DVH+ADV model had the lowest prediction error of 0.038 (conformation), 0.026 (homogeneity), 0.298 (R50), 1.65% (D95), 2.14% (D98), 2.43% (D99). The MSE model had the worst prediction error of 0.134 (conformation), 0.041 (homogeneity), 0.520 (R50), 3.91% (D95), 4.33% (D98), 4.60% (D99). For both the mean dose PTV error and the max dose PTV, Body, Bladder and rectum error, the MSE+DVH+ADV outperformed all other models. Regardless of model, all predictions have an average mean and max dose error less than 2.8% and 4.2%, respectively.

**Conclusion**: The MSE+DVH+ADV model performed the best in these categories, illustrating the importance of both human and learned domain knowledge. Expert human domain specific knowledge can be the largest driver in the performance improvement, and adversarial learning can be used to further capture nuanced attributes in the data. The real-time prediction capabilities allow for a physician to quickly navigate the tradeoff space for a patient, and produce a dose distribution as a tangible endpoint for the dosimetrist to use for planning. This is expected to considerably reduce the treatment planning time, allowing for clinicians to focus their efforts on the difficult and demanding cases.

# I. Introduction

External beam radiation therapy is one of the major treatments available to cancer patients, with major modalities available including intensity modulated radiation therapy (IMRT)[1-7] and volume modulated arc therapy (VMAT)[8-15]. IMRT and VMAT have revolutionized the treatment planning over the past decades, drastically improving the treatment plan quality and patient outcome. However, many tedious and time consuming aspects still exist within the clinical treatment planning workflow. In particular, there are two aspects: 1) The dosimetrist must tediously and iteratively tune the treatment planning hyperparameters of the fluence map optimization in order to arrive at a planner-acceptable dose, and 2) many feedback loops between the physician and dosimetrist occur for the physician to provide his comments and judgement on the plan quality, until a physician-acceptable dose is finally produced. For a patient, this process can continually repeat for many hours to many days, depending on the complexity of the plan.

Much work over the years has been focused on reducing the treatment complexity by simplifying certain aspects in the planning workflow, such as feasibility seeking[16], multicriteria optimization for tradeoff navigation on the Pareto surface[17-19], and other algorithms for performance improvements[20-26]. While effective, these methods still require a large amount of intelligent input from the dosimetrist and physician, such as in weight tuning and deciding appropriate dose-volume constraints and tradeoffs.

To address this, the field developed machine learning models to predict relevant dosimetric endpoints, which can be categorized into 1 of 3 categories: 1) predicting single dose constraint points directly, 2) predicting dose volume histograms (DVH), and 3) predicting the 3D dose distribution of the plan. With a 3D dose distribution, one can fully reconstruct the DVH, and with the DVH, the dose constraints can then be calculated. Many studies focused on either predicting dose constraints or the dose volume histogram, eventually forming the backbone of knowledge-based planning (KBP)[27-42]. KBP used machine learning techniques and models to predict clinically acceptable dosimetric criteria, utilizing a large pool of historical patient plans and information to draw its knowledge from. Before the era of deep neural networks, KBP's efficacy was heavily reliant on not only the patient data size and diversity, but also on the careful selection of features extracted from the data to be used in the model[32-39,42,43]. This limited the model to predict small dimensional data, such as the DVH or specific dosimetrist criteria.

With the advancements in deep learning, particularly in computer vision[44-46] and convolutional neural networks[47], many studies have investigated clinical dose distribution prediction using deep learning on several sites such as for prostate IMRT[48,49], prostate VMAT[32], lung IMRT[50], head-and-neck IMRT[51-54], head-and-neck VMAT[55]. In addition to clinical dose prediction, deep learning models are capable of accurately generating Pareto optimal dose distributions, navigating the various tradeoffs between planning target volume (PTV) dose coverage and organs-at-risk (OAR) dose sparing[56]. Most of these methods utilize a simple loss function for training the neural network—the mean squared error (MSE) loss. MSE loss is a generalized, domain-agnostic loss function that can be applied to many problems in many domains. It's large flexibility also means that it is incapable of driving its performance in a domain-specific manner.

Mahmood and Babier et al.[52-54] investigated the use of adversarial learning for dose prediction. Since the development of generative adversarial networks (GAN) by Goodfellow[57], adversarial loss has been popularized in the deep learning community for many applications. While used heavily for generative models, such as GANs, the adversarial loss can be applied to almost any neural network training. The adversarial loss's emerging success in deep learning application is largely due to the discriminator capability to calculate its own feature maps during the training process. In essence, the discriminator is learning its own domain knowledge of the problem. However, an adversarial framework is not without its issues. The user has little control over what kinds of features the discriminator may be learning during the training process. It is possible for the discriminator to learn the correct answer for the wrong reason. In addition, careful balancing of the learning between the two networks is essential for preventing catastrophic failure. These may affect the overall performance of the prediction framework.

In 2018, Muralidhar et al.[58] proposed a domain adapted loss into their neural network training, in order to address deep learning in cases of limited and poor-quality data, which is a problem commonly found within the medical field. They found that, by including domain-explicit constraints, the domain adapted network model had drastically improved performance over its domain-agnostic counterpart, especially in the limited, poor-quality data situation. We realize the importance of including domain specific losses into the radiation therapy problem of dose prediction. We propose the addition of a differentiable loss function based on the dose volume histogram (DVH), one of the most important and commonly used metrics in radiation oncology, into the training of deep neural networks for volumetric dose distribution prediction. In this study, we will train a neural network for generating Pareto optimal dose distributions, and evaluate the effects of MSE loss, DVH loss, and adversarial loss on the network's performance. Pareto optimal plans are the solutions to a multicriteria problem with various tradeoffs. In particular, the tradeoff lies with the dose coverage of the tumor and the dose sparing of the various critical structures. The benefit of such a model is two-fold. First, the physician can interact with the model to immediately view a dose distribution, and then adjust some parameters to push the dose towards their desired tradeoff in real time. This also allows for the physician to quickly comprehend the kinds of the tradeoffs that are feasible for the patient. Second, the treatment planner, upon receiving the physician's desired dose distribution, can quickly generate a fully deliverable plan that matches this dose distribution, saving time in tuning the optimization hyperparameters and discussing with the physician.

Because generating Pareto optimal plans for the patient requires for the network to learn how to map many dose distributions with tradeoffs from a single anatomy, the neural network must learn to differentiate the potentially small nuances between the different doses that may have substantial clinical consequences. While these small nuances may not be well reflected in a metric such as voxel-wise mean squared error, our own domain metrics can amplify the clinically relevant differences. The usage of an adversarial loss can further aid the neural network in learning important differences that cannot be easily formulated into a loss function. We believe that training a network to generate Pareto optimal dose distribution is well suited for testing the effects of MSE loss, DVH loss, and adversarial loss.

## II. Methods

### II.1. Patient and Pareto Optimal Plan Data

The data for 70 prostate patients, including the planning target volume (PTV), and the organs-at-risk (OAR)—body, bladder, rectum, left femoral head, and right femoral head—were acquired as 96 x 96 x 24 dimension arrays at 5 mm³ voxel size. Ring and skin structures were added as tuning structures. The dose influence arrays were calculated for the 70 patients, using a 7 equidistant coplanar beam plan IMRT setup. The beamlet size was 2.5 mm² at 100 cm isocenter. Using this dose influence data, we generated IMRT plans that sampled the Pareto surface, representing various tradeoffs between the PTV dose coverage and OAR dose sparing. The multicriteria objective can be written as:

$$\underset{x}{argmin} \quad \{f_{PTV}(x), f_{OAR_1}(x), f_{OAR_2}(x), \ldots, f_{OAR_{n_{OAR}}}(x)\}$$
$$\text{subject to} \quad x \geq 0$$

(1)

where $x$ is the fluence map intensities to be optimized. The individual objectives, $f_s(x) \; \forall s \in \{PTV, OAR_u \; \forall u \in \{1,2,\ldots,n_{OAR}\}\}$, are for the PTV and each of the OARs used in the optimization problem, where $n_{OAR}$ represents the total number of OARs. The index $s$ represents a structure used in the optimization, which is the PTV or one of the OARs. For simplicity, we define $S$ as the set of all structures used in the optimization. In this case, $S = \{PTV, OAR_u \; \forall u \in \{1,2,\ldots,n_{OAR}\}\}$ and $s \in S$. In radiation therapy, the objective function is formulated with the intention to deliver the prescribed dose to the PTV, while minimizing the dose to each OAR. Because to the physical aspects of radiation in external beam radiation therapy, it is impossible to deliver to the PTV the prescription dose without irradiating normal tissue. In addition, it has been shown that the integral dose to the body is similar regardless of the plan[59-62], so, in essence, one can only choose how to best distribute the radiation in the normal tissue. For example, by reducing the dose to one OAR, either the PTV coverage will worsen or more dose will be delivered to another OAR. Therefore, we arrive at a multicriteria objective, where there does not exist a single optimal $x^*$ that would minimize all $f_s(x) \; \forall s \in PTV, OAR$. In this study, we choose to use the $\ell_2$-norm to represent the objective, $f_s(x) = \frac{1}{2}\|A_s x - p_s\|_2^2$. In this formulation, $A_s$ is the dose influence matrix for a given structure, and $p_s$ is the desired dose for a given structure, assigned as the prescription dose if $s$ is the PTV, and 0 otherwise. Our beamlet-based dose influence matrix was calculated using the Eclipse AAA dose calculation engine (Varian Medical Systems, Inc.). This allows for us to linearly scalarize the multicriteria optimization problem[63], by reformulating it into a single-objective, convex optimization problem:

$$\underset{x}{argmin} \quad \frac{1}{2}\sum_{s \in S} w_s^2 \|A_s x - p_s\|_2^2$$
$$\text{subject to} \quad x \geq 0$$

(2)

Scalarizing the problem required the addition of new hyperparameters, $w_s$, which are the tradeoff weights for each objective function, $f_s(x) \ \forall s \in S$. By varying $w_s$ to different values, different Pareto optimal solutions can generated by solving the optimization problem. Using an in-house GPU-based proximal-class first-order primal-dual algorithm, Chambolle-Pock[64], we generated many pseudo-random plans, by assigning pseudo-random weights to the organs-at-risk. The weight assignment fell into 1 of 6 categories as shown in Table 1.

| Category | | Description |
|---|---|---|
| Single organ spare | Bladder | $w_{bladder} = rand(0,1)$<br>$w_{OAR\backslash bladder} = rand(0,0.1)$ |
| | Rectum | $w_{rectum} = rand(0,1)$<br>$w_{OAR\backslash rectum} = rand(0,0.1)$ |
| | Lt Fem Head | $w_{lt\ fem\ head} = rand(0,1)$<br>$w_{OAR\backslash lt\ fem\ head} = rand(0,0.1)$ |
| | Rt Fem Head | $w_{rt\ fem\ head} = rand(0,1)$<br>$w_{OAR\backslash rt\ fem\ head} = rand(0,0.1)$ |
| | Shell | $w_{shell} = rand(0,1)$<br>$w_{OAR\backslash shell} = rand(0,0.1)$ |
| | Skin | $w_{skin} = rand(0,1)$<br>$w_{OAR\backslash skin} = rand(0,0.1)$ |
| High weights | | $w_s = rand(0,1) \ \forall s \in OAR$ |
| Medium weights | | $w_s = rand(0,0.5) \ \forall s \in OAR$ |
| Low weights | | $w_s = rand(0,0.1) \ \forall s \in OAR$ |
| Extra low weights | | $w_s = rand(0,0.05) \ \forall s \in OAR$ |
| Controlled weights | | $w_{bladder} = rand(0,0.2)$<br>$w_{rectum} = rand(0,0.2)$<br>$w_{lt\ fem\ head} = rand(0,0.1)$<br>$w_{rt\ fem\ head} = rand(0,0.1)$<br>$w_{shell} = rand(0,0.1)$<br>$w_{skin} = rand(0,0.3)$ |

Table 1: Weight assignment categories for the organs at risk. The function $rand(lb, ub)$ represents a uniform random number between a lower bound, $lb$, and an upper bound, $ub$. In the high, medium, low, extra low, and controlled weights category, the PTV had a 0.05 probability of being assigned $rand(0,1)$ instead of 1.

For each patient, 100 plans of the single organ spare category (bladder, rectum, left femoral head, right femoral head, shell, skin) were generated for each critical structure, yielding 600 organ sparing plans per patient. To further sample the tradeoff space, another 100 plans of the high, medium, low, and extra low weights category were generated, as well as 200 plans of the controlled weights category. In the high, medium, low, extra low, and controlled weights category, the PTV had a 0.05 probability of being assigned $rand(0,1)$ instead of 1. The bounds for the controlled weights were selected through trial-and-error such that the final plan generated was likely to fall within clinically relevant bounds, although it is not necessarily acceptable by a physician. In total 1200 plans were generated per patient. With 70 patients, the total number of plans generated was 84,000 plans.

Regarding time for data generation, for each patient, on average it takes 32 minutes to use the Eclipse AAA engine to compute beamlet-based dose influence matrices for a 5 beam IMRT plan. Using our GPU-accelerated optimization algorithm, it takes roughly 2 seconds to generate 1 Pareto optimal plan. These exclude the time it takes to identify and gather the original patient data, as well as preprocessing steps required for converting the data into python-friendly formats. While the optimization of the Pareto optimal plans is fast, the bottleneck is the dose influence matrix calculation for each patient. This is an additional motivation for using neural networks, which do not require dose influence matrices for predicting Pareto optimal dose distributions. This has been shown to yield substantial time savings in generating Pareto optimal plans[56].

## II.2. Loss Functions

In this study, 3 loss functions—mean squared error (MSE) loss, dose volume histogram (DVH) loss, and adversarial (ADV) loss—were used to train and compare 4 instances of the neural network model. The first model used only the voxel-wise MSE loss. The second model's loss function used the MSE loss in conjunction with the ADV loss. The third model used the MSE loss in conjunction with the DVH loss. The fourth and last model's loss function combined MSE, DVH, and ADV losses all together. Respectively, the study will denote each variation as MSE, MSE+ADV, MSE+DVH, and MSE+DVH+ADV. The following section will describe the ADV and DVH losses in detail.

## II.1.1. Adversarial Loss

Our adversarial-style training utilizes a framework similar to that of generative adversarial networks (GAN)[57], with respect to having another model acting as a discriminator to guide the main network to produce a dose distribution close to the real data. The major benefit to this approach is that the discriminator is calculating its own features and metrics to distinguish the ground truth data and predicted data. Effectively, this is allowing the discriminator to learn its own domain knowledge, and then provide feedback to update the main model. For this study, we utilized the Least Squares GAN (LSGAN)[65] formulation:

$$\operatorname*{argmin}_{\theta_{N_D}} L_{ADV_D} = \frac{1}{2}\|N_D(y_{true}) - b\|_2^2 + \frac{1}{2}\|N_D(N_G(x)) - a\|_2^2$$

(3)

$$\operatorname*{argmin}_{\theta_{N_G}} L_{ADV_G} = \frac{1}{2}\|N_D(N_G(x)) - c\|_2^2$$

(4)

where $\theta_{N_D}$ and $\theta_{N_G}$ are the trainable weights parameterizing the discriminator, $N_D$, and generator, $N_G$, respectively. $L_{ADV_D}$ and $L_{ADV_G}$ are the loss functions to be minimized with respect to $\theta_{N_D}$ and $\theta_{N_G}$. The variable $x$ represents the input into the generator, $N_G$, which, because of $L_{ADV_G}$, tries to create data that has a similar distribution to that of $y_{true}$, our target data. The discriminator tries to distinguish $y_{true}$ from the data created from the generator. As per suggestion by the LSGAN publication[65], to minimize the Pearson $X^2$ divergence, we set $a = -1$, $b = 1$, and $c = 0$.

### II.1.2. Dose Volume Histogram Loss

The DVH is one of the most commonly used metrics in radiation oncology for evaluating the quality of a plan, so it is natural to assume that including this metric as part of the loss would be beneficial. However, the calculation of the DVH involves non-differentiable operations, which means any loss based on it cannot provide a gradient to update the neural network. We propose a differential approximation of the DVH, which we define as $\widetilde{DVH}$. Given a binary segmentation mask, $M_s$, for the $s^{th}$ structure, and a volumetric dose distribution, $D$, the volume at or above a given dose threshold value, $d_t$, can be approximated as:

$$v_{s,d_t}(D, M_s) = \frac{\sum_{i,j,k} Sigmoid\left(\frac{m}{\beta_t}(D(i,j,k) - d_t)\right) M_s(i,j,k)}{\sum_{i,j,k} M_s(i,j,k)}$$

(5)

where $Sigmoid(x) = \frac{1}{1+e^{-x}}$ is the sigmoid function, $m$ controls the steepness of the curve, $\beta_t$ is the bin width of the histogram, and $i, j,$ and $k$ are the voxel indices for the 3D arrays. The $t$ is an index for the dose threshold values and bin widths. The total number of thresholds is defined as $n_t$, which is also equivalent to the number of bins in $\widetilde{DVH}$. We also constrain the dose to be monotonically increasing with increasing index, $d_1 \leq d_2 \leq \ldots \leq d_{n_t}$. Based on this, the $\widetilde{DVH}_s$ for any structure, $s$, can then be defined as:

$$\widetilde{DVH}_s(D, M_s) = \left(v_{s,d_1}, v_{s,d_2}, \cdots, v_{s,d_{n_t}}\right)$$

(6)

The bin centers and widths are then defined as:

$$center_{bin} = \left(\frac{d_1 + d_2}{2}, \frac{d_2 + d_3}{2}, \ldots, \frac{d_{n_t} + d_{n_t+1}}{2}\right)$$

(7)

$$width_{bin} = (\beta_1, \beta_2, \cdots, \beta_{n_t}) = (d_2 - d_1, d_3 - d_2, \cdots, d_{n_t+1} - d_{n_t})$$

(8)

To illustrate Equation 3, we calculated the DVH and the approximated DVH, of varying steepness values of $m = \{1,2,4,8\}$, of a PTV and OAR or an example prostate patient. As demonstrated by Figure 1, when the steepness of the curve $m \to \infty$, then $\widetilde{DVH} \to DVH$.

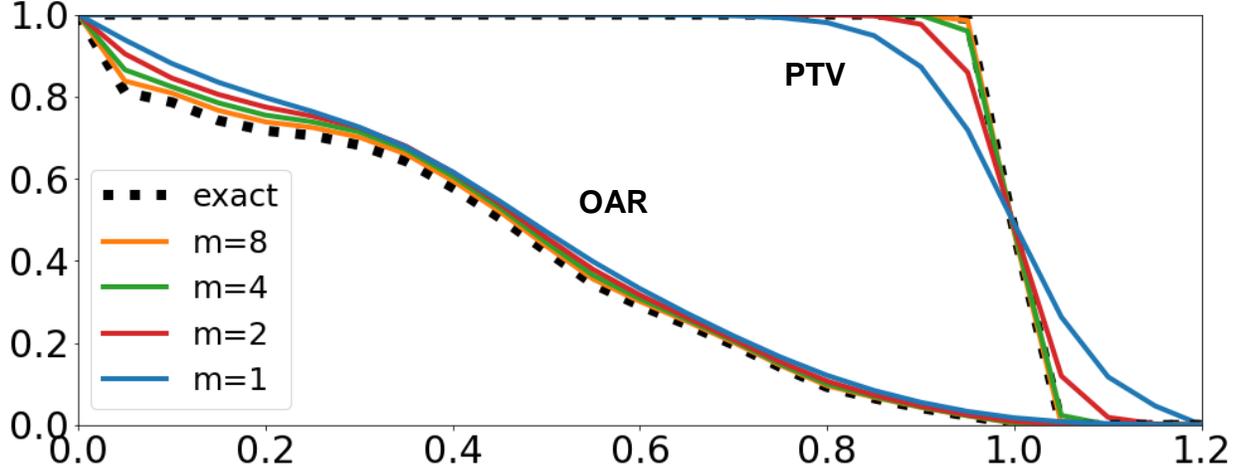

Figure 1: DVH and approximated DVH calculations using varying steepness values of $m = \{1,2,4,8\}$ for an example prostate patient.

Because $\widetilde{DVH}$ is computed using sigmoid, the gradient, $\frac{\partial \widetilde{DVH}(D,M)}{\partial D} = \left(\frac{v_{s,d_0}}{\partial D}, \frac{v_{s,d_1}}{\partial D}, \cdots, \frac{v_{s,d_n}}{\partial D}\right)$, can be computed, allowing for a loss function utilizing $\widetilde{DVH}$ to be used to update the neural network weights. We can then define a mean squared loss of the DVH as:

$$L_{DVH}(D_{true}, D_{pred}, M) = \frac{1}{n_s}\frac{1}{n_t}\sum_s \left\|\widetilde{DVH}_s(D_{true}, M_s) - \widetilde{DVH}_s(D_{pred}, M_s)\right\|_2^2$$

(9)

where $D_{true}$ and $D_{pred}$ are the ground truth and predicted doses, respectively. While a gradient of $L_{DVH}$ exists, it is possible that the gradient space is ill-behaved and would be not suitable for use. We studied the properties of this approximation using a simple toy example. Letting $D = (1,2)$, The exact DVH and approximate DVH with varying values of $m = \{1,2,4,8\}$ can be calculated, shown in Figure 2.

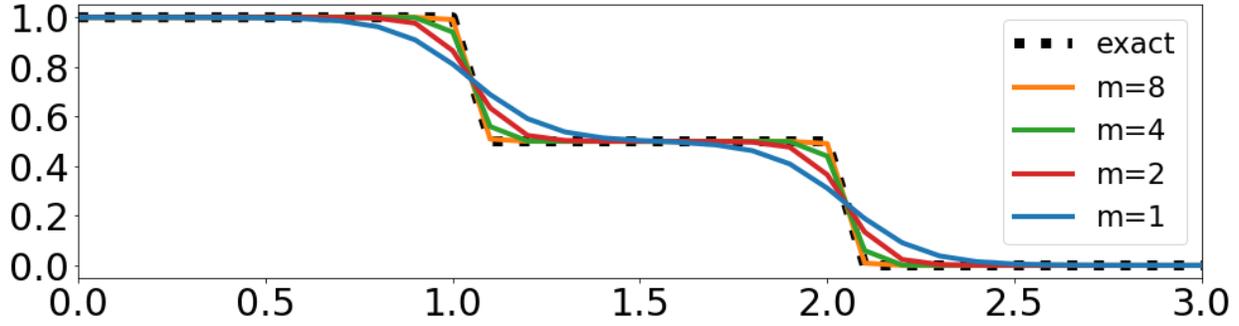

Figure 2: DVH and approximated DVH calculations of toy example for $D = (1,2)$.

It can be observed that in this toy example in Figure 2 the approximation is smoother and has larger error with smaller $m$, which agrees with Figure 1. To investigate the gradient properties of the loss using the approximate DVH, we calculated $\left\|DVH([1,2], M) - \widetilde{DVH}([i,j], M)\right\|_2^2 \forall i,j \in (0,3)$ for $M = [1,1]$.

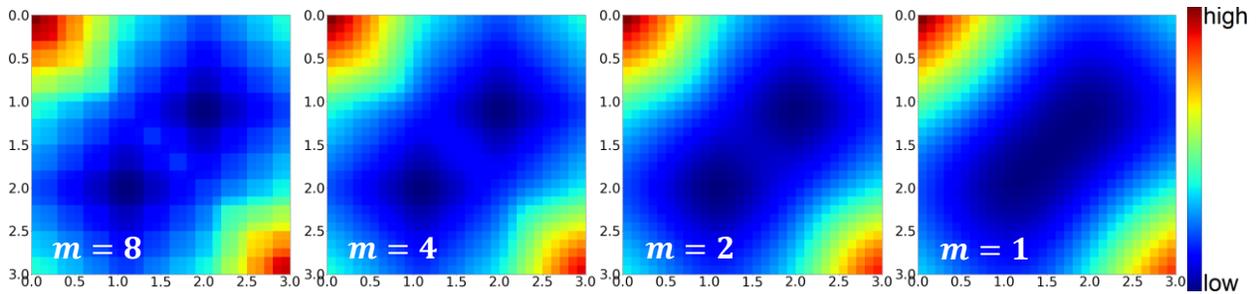

Figure 3: Objective value map of the loss function $\left\|DVH([1,2], M) - \widetilde{DVH}([i,j], M)\right\|_2^2 \forall i,j \in (0,3)$ for $M = [1,1]$. All versions with varying values of $m$ exhibit the same minima.

Figure 3 shows the squared error value of the difference between DVH for the data $(1,2)$ and the $\widetilde{DVH}$ for the data $(i,j) \ \forall i,j \in (0,3)$. There are multiple local minima. For our case it is when $(i,j) = (1,2) \ or \ (2,1)$, since either will produce the same DVH. For higher $m$, the local minimas become more defined, with steeper gradients surrounding them, an undesirable quality for optimization and learning problems. While a lower steepness, $m$, may not approximate the DVH as well, the loss function involving the $\widetilde{DVH}$ maintains the same local minima, and provides a smoother, and more well-behaved gradient than its sharper counterparts. For the remainder of this study, we chose to use $\widetilde{DVH}$ with $m = 1$.

## II.3. Model Architecture

In this study the dose prediction model that was utilized was a U-net style architecture[66], and the discriminator model was a classification style architecture[67,68].

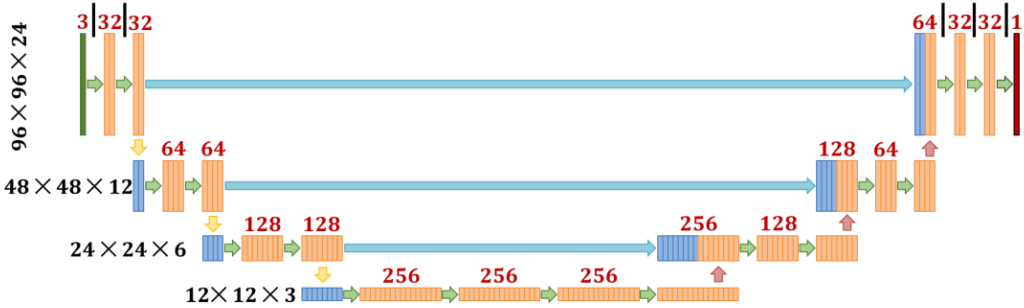

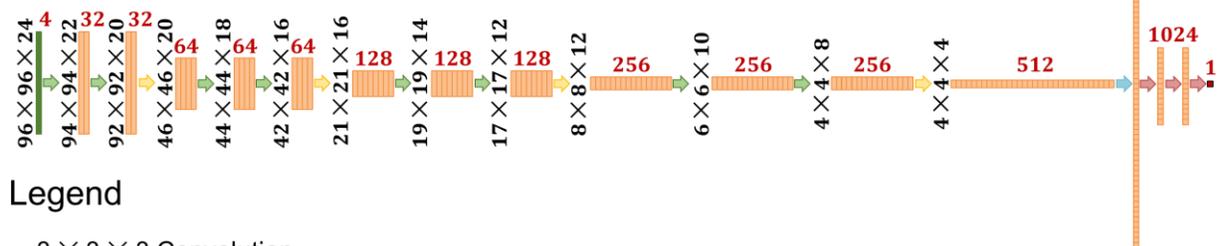

Figure 4: Deep learning models used in the study. The same U-net architecture is utilized in each comparison of MSE, MSE+ADV, MSE+DVH, and MSE+DVH+ADV models. The same discriminator architecture is utilized for training the MSE+ADV and MSE+DVH+ADV models. Black numbers to the left of the feature blocks represent the current data shape. Red numbers above the feature blocks represents the number of features.

Specifically, the models were adjusted to match the data shape. The architectures shown in Figure 4 depict the models used in the study. The U-net takes as input a 3 channel tensor that consists of, 1) the PTV mask with the value $w_{PTV}$ assigned as its non-zero value, 2) the OAR masks that include all the OARs respectively assigned their $w_{OARs}$, and 3) the body mask as a binary. The U-net then performs multiple calculations, with max pooling operations to reduce the resolution for more global feature calculation, and then upsampling operations to eventually restore the image resolution back to the original. Concatenation operations are used to merge the local features calculated in the first half of the U-net with the global features calculated at the bottom and later half of the U-net.

The discriminator architecture is of an image classification style network. The goal of the discriminator is to learn how to distinguish the optimized dose distributions versus the U-net predicted dose distribution. Similar to conditional generative adversarial network framework[69], the discriminator will additionally take the same input as the U-net. In total, the input data has 4 channels—3 channels of the U-net's input and 1 channel of either the optimized or predicted dose distribution. As shown in Figure 4, the discriminator goes through a process of convolutions and strided convolutions to calculate new feature maps and to reduce the image resolution, respectively. It is important to note that the strided convolution is used to reduce one or more of the image dimensions by half, but differ in which dimensions are being reduced in order to eventually reduce the image to 4 x 4 x4 pixels. For example, the first strided convolution is applied to the first 2 image dimensions, reducing the image from 92 x 92 x 20 to 46 x 46 x20, but the last strided convolution is reducing the 3rd image dimension. The specific details can be seen in the image sizes specified in Figure 4.

In addition, Group Normalization[70] was used in place of Batch Normalization[71], which has been shown to allow for the models to effectively train on small batch sizes. All activations in the hidden layer are rectified linear units (ReLU) activations. Final activations for both the U-net and discriminator are linear activations.

## II.4. Training and Evaluation

We first notate the mean squared error loss, dose volume histogram loss, and U-net's adversarial loss as $L_{MSE}(y_t, y_p)$, $L_{DVH}(y_t, y_p, M)$, and $L_{ADV_G}(y_t, y_p, x)$, where $x$ is the input into the U-net model, $y_t$ is the ground truth optimized dose distribution, $y_p$ is the predicted dose distribution, and $M$ contains the binary segmentation masks. The total objective for training the U-net is then defined as:

$$L_{Total} = L_{MSE}(y_t, y_p) + \lambda_{DVH} L_{DVH}(y_t, y_p, M) + \lambda_{ADV} L_{ADV_G}(y_p, x),$$

(10)

and the objective for training the discriminator is simply $L_{ADV_D}(y, x)$, where $y$ can either be $y_t$ or $y_p$ for a given training sample. For each study—MSE, MSE+ADV, MSE+DVH, and MSE+DVH+ADV—the weightings, $\lambda_{DVH}$ and $\lambda_{ADV}$, used are shown in Table 2. These were chosen by evaluating the order of magnitude of the values that each loss function exhibits for a converged model. From previous dose prediction studies and results[48,55], we can estimate that the $L_{MSE} \sim 10^{-4}$ and $L_{DVH} \sim 10^{-3}$ for a converged model. Since we are using least squares GAN framework, we estimate the loss $L_{ADV_G}$ ranges from $10^{-1}$ to $10^0$. Our choice of $\lambda_{DVH}$ and $\lambda_{ADV}$, shown in Table 2, is to have each component of the loss to be within a similar order of magnitude for when the model is converged.

|  | $\lambda_{DVH}$ | $\lambda_{ADV}$ |
|---|---|---|
| MSE | 0 | 0 |
| MSE+ADV | 0 | 0.001 |
| MSE+DVH | 0.1 | 0 |
| MSE+DVH+ADV | 0.1 | 0.001 |

Table 2: Choices of $\lambda_{DVH}$ and $\lambda_{ADV}$ for the loss function shown in Equation 10.

We divided the 70 prostate patients into 54 training, 6 validation, and 10 testing patients, yielding 64,800 training, 7,200 validation, and 12,000 testing plans. For the training that involved adversarial loss, the U-net and discriminator would alternate every 100 iterations, to allow for some stability in the training and loss. The discriminator was trained to take as input the same inputs as the u-net as well as a dose distribution, either from the real training data or from the U-net's prediction. With a 0.5 probability, the discriminator would receive either real training data or data predicted from the U-net. Each U-net model was trained for a total of 100,000 iterations, using a batch size of 2. All models used the Adam optimizer, with a learning rate of $1 \times 10^{-3}$. All training was performed on an NVIDIA 1080 Ti GPU with 11 GB RAM. After training, the model with the lowest total validation loss was used to assess the test data.

All dose statistics will also be reported relative to the relative prescription dose (i.e. the errors are reported as a percent of the prescription dose). As clinical evaluation criteria PTV coverage (D98, D99), PTV max dose, homogeneity $\left(\frac{D2-D98}{D50}\right)$, van't Riet conformation number[72] $\left(\frac{(V_{PTV} \cap V_{100\%Iso})^2}{V_{PTV} \times V_{100\%Iso}}\right)$, the dose spillage $R50$ $\left(\frac{V_{50\%Iso}}{V_{PTV}}\right)$, and the structure max and mean doses ($D_{max}$ and $D_{mean}$) were evaluated. $D_{max}$ is defined as the dose to 2% of the structure volume, as recommended by the ICRU report no 83[73].

## III. Results

For each model, training for 100,000 iterations took 1.5 days (MSE), 3.5 days (MSE+ADV), 2.3 days (MSE+DVH), 3.8 days (MSE+DVH+ADV). After training, the prediction time of each U-net is the same at 0.052 seconds, since all 4 U-net models in the study are identical in architecture.

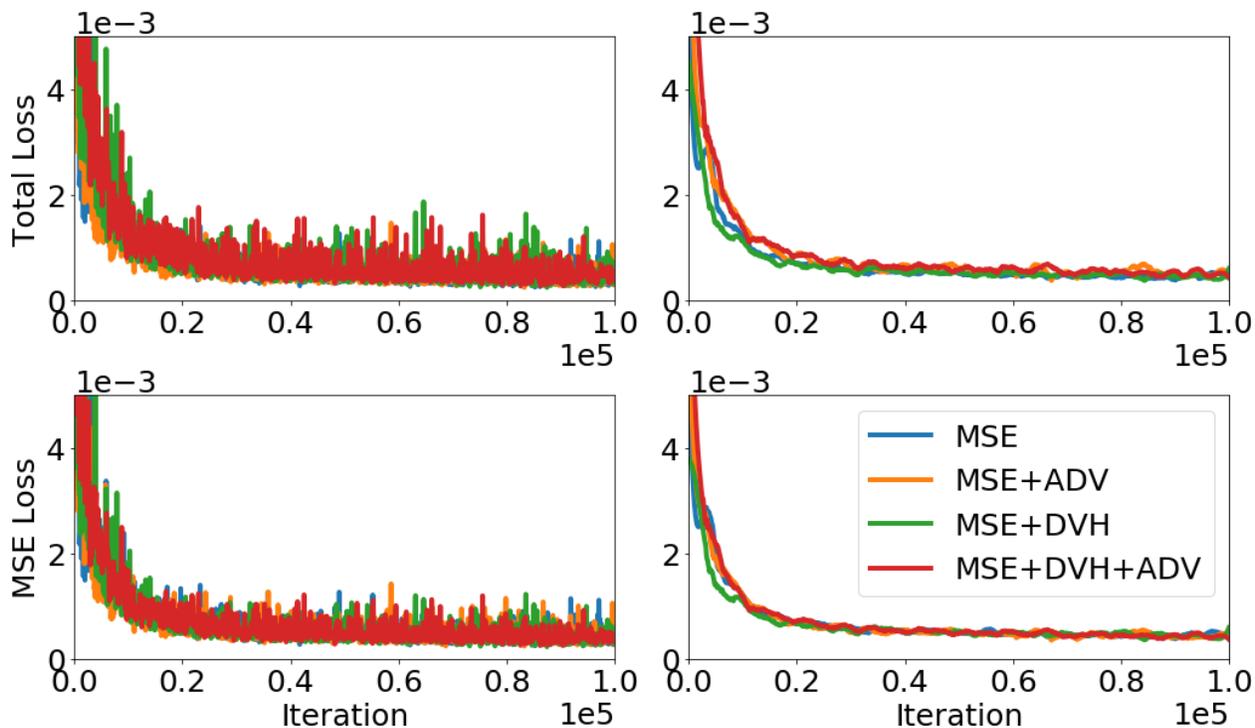

Figure 5: Top row: Total validation loss (all relevant losses and loss weightings for a specific model are summed are displayed). Bottom row: MSE validation loss (only mean squared error is displayed). Left Column: Raw validation losses at each training iteration. Right Column: Smoothed validation loss using Savitzky–Golay filter[74].

Figure 5 shows the validation losses for each model. The top row shows the total validation loss, while the bottom row shows just the mean squared error loss. Overall the loss curve had flattened by the end of the 100,000 iterations, indicating that each model converged. The final instances of the models chosen for evaluation were the models that performed the best on their respective total validation loss. Each model has achieved similar MSE losses, with our chosen models having their MSE validation losses at $2.46 \times 10^{-4}$ (MSE), $2.40 \times 10^{-4}$ (MSE+ADV), $2.26 \times 10^{-4}$ (MSE+DVH), $2.5 \times 10^{-4}$ (MSE+DVH+ADV).

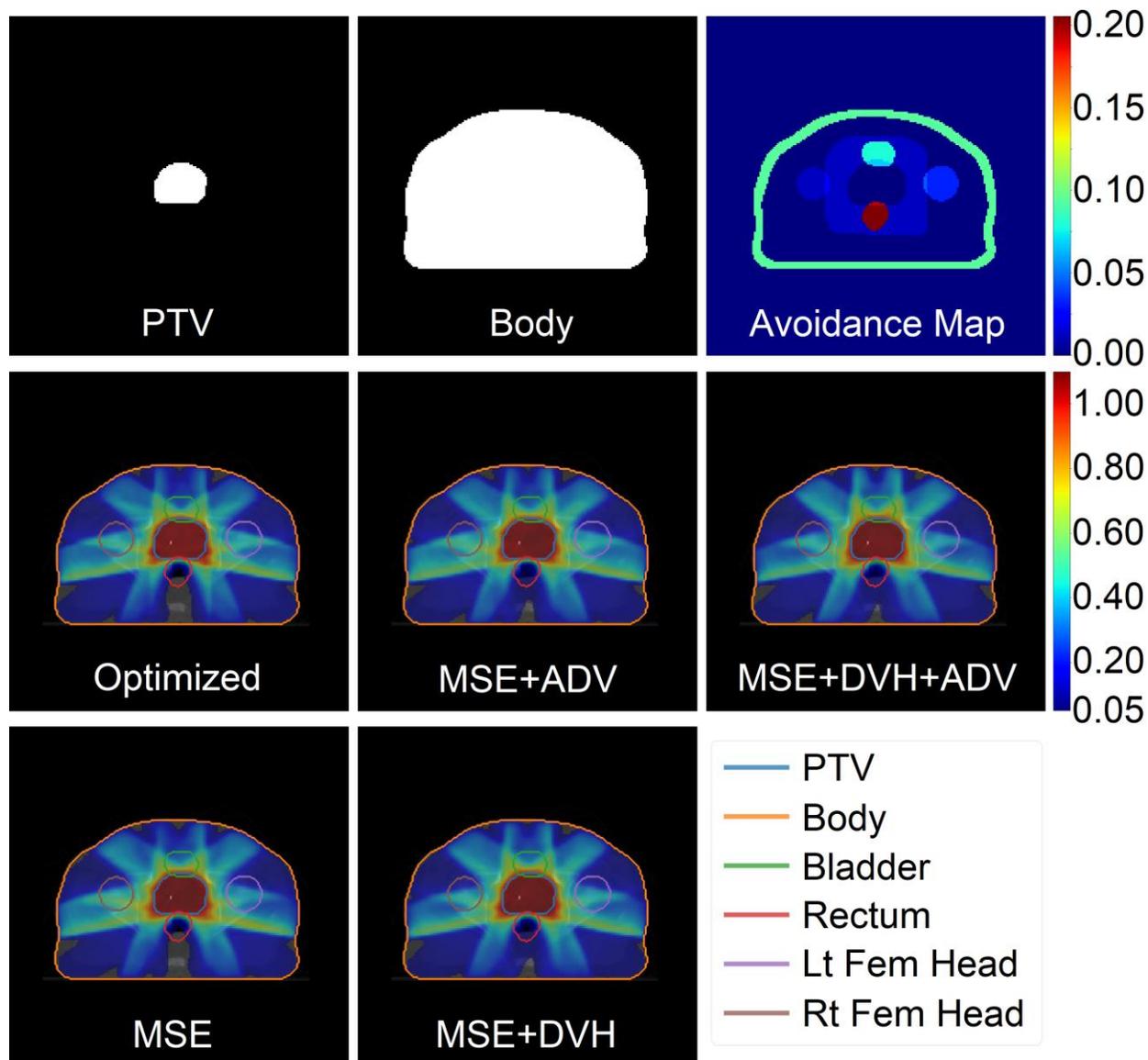

Figure 6: Inputs, optimized dose, and predicted doses for a test patient and a rectum sparing plan. Top row: Inputs of the U-net neural network, which include the PTV assigned to its weight ($w_{PTV} = 1$ in this example), a binary mask of the body, and the avoidance map containing the remaining organs-at-risk assigned to their respective tradeoff weight. Bottom two rows: Optimized and predicted dose washes of the Pareto optimal dose. The colorbar shows the doses between 5% and 110% of the prescription dose.

Figure 6 shows a comparison of the predictions between each of the 4 models in the study on 1 example patient and Pareto optimal plan. The "optimized" dose is the ground truth Pareto optimal dose that was obtained by solving the optimization problem outlined in Equations 1 and 2. The avoidance map is a sum of the critical structures, including a ring and skin tuning structure, with their assigned tradeoff weights. The 4 models each take in the top row of Figure 6 as their input, and then predict what the Pareto optimal plan should look like. Visually, with the same input, each model produces a strikingly similar dose distribution to the optimized case. The MSE model visually did slightly better in sparing the dose in the normal tissue region posterior of the rectum.

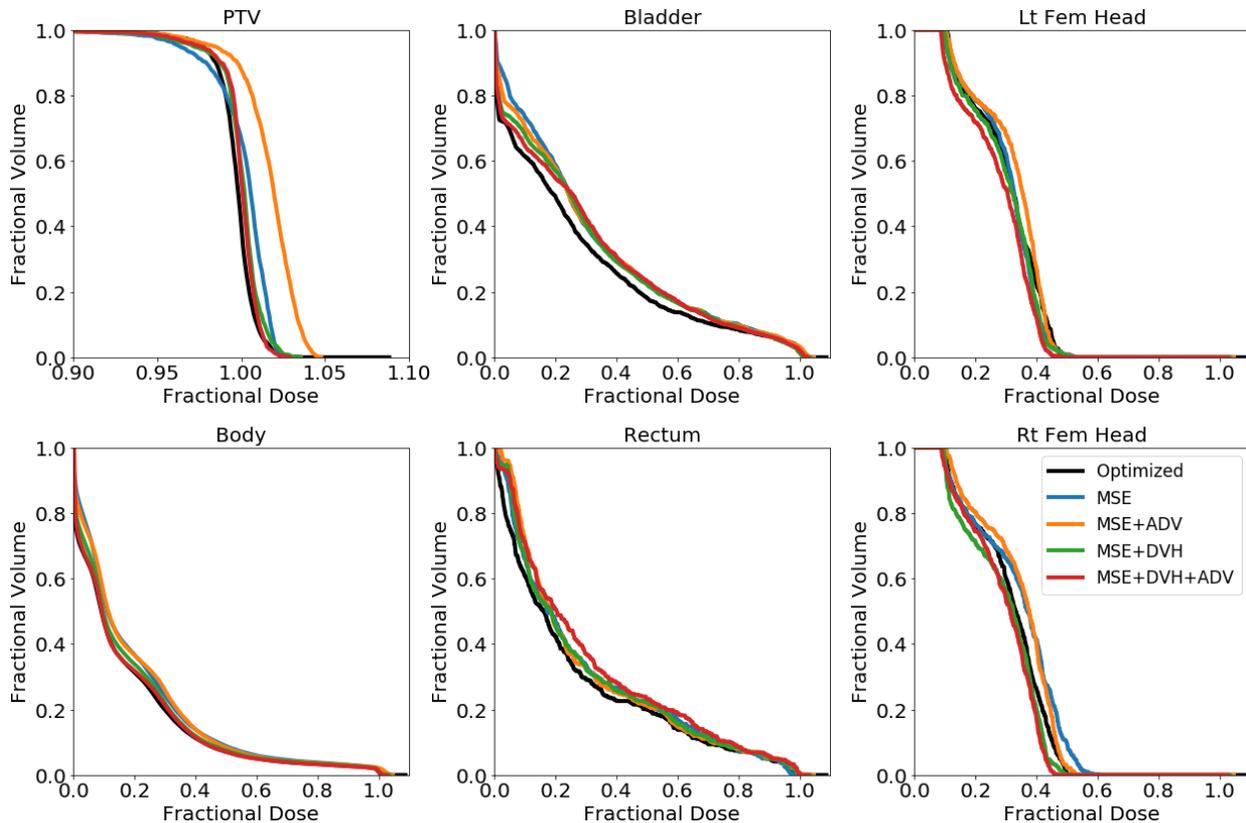

Figure 7: Dose volume histograms (DVH) of optimized dose distribution (black) and predicted dose distributions (various colors) for the same test patient as in Figure 6. Note the x-axis scale for the PTV DVH is different.

The DVHs of the dose predictions are more revealing to the dose prediction errors in a clinically relevant manner, shown in Figure 7. The two models involving DVH loss (red and green) have less error in predicting the dose in the PTV, Body, and Bladder, with respect to its DVH, and visually similar predictions for the remaining OARs. Overall, including the domain specific DVH based loss has overall improved the model's dose prediction in regards to the structure's DVH.

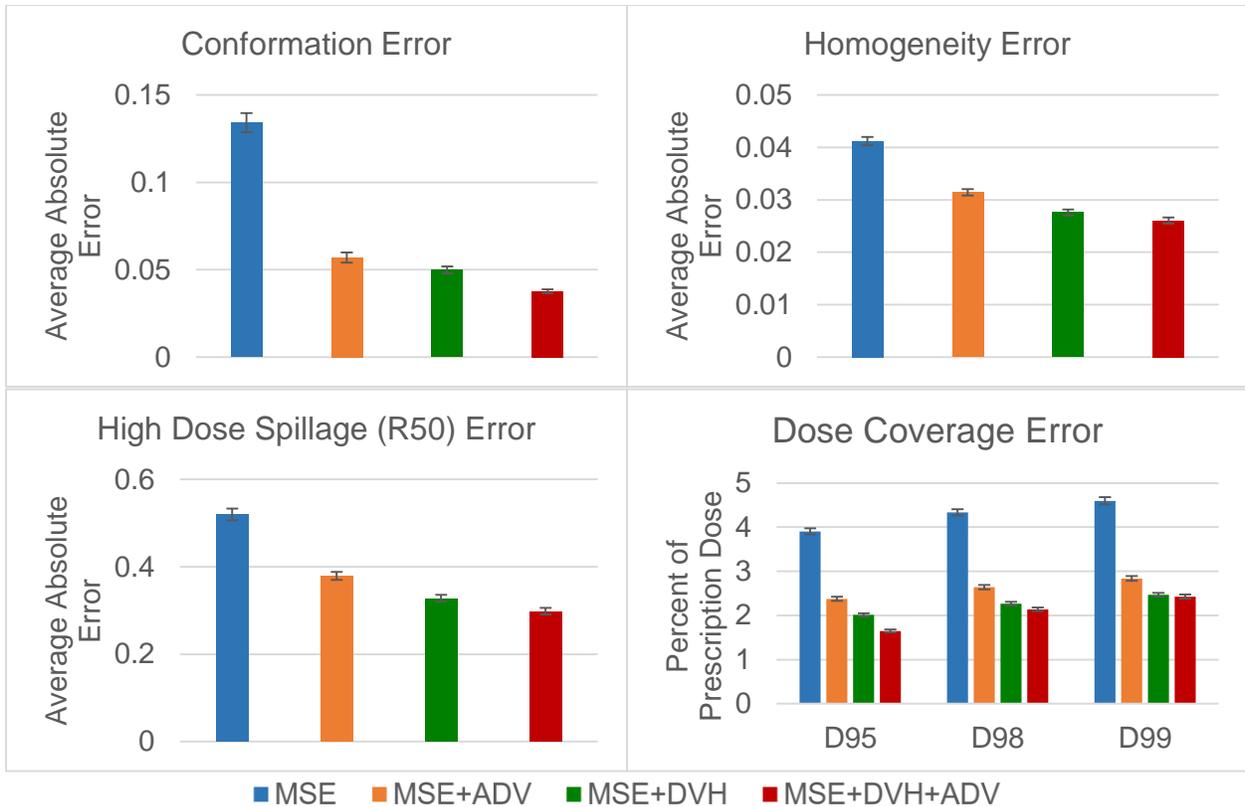

Figure 8: Predicion errors for conformation, homogeneity, high dose spillage (R50) and dose coverage on the test data. Error bar represents the 99% confidence interval $\left(\bar{x} \pm 2.576 * \frac{\sigma}{\sqrt{n}}\right)$, taken overall all test patients and plans.

Figure 8 shows the errors for several clinical metrics calculated from the predicted dose distributions, as compared to that of the optimized dose. The MSE model had the largest prediction error of 0.134 (conformation), 0.041 (homogeneity), 0.520 (R50), 3.91% (D95), 4.33% (D98), 4.60% (D99). The additional adversarial and DVH losses further improved the prediction error, with the MSE+DVH+ADV model having the lowest prediction error of 0.038 (conformation), 0.026 (homogeneity), 0.298 (R50), 1.65% (D95), 2.14% (D98), 2.43% (D99). The other two models, MSE+ADV and MSE+DVH, had errors that were between the other two, with the MSE+DVH model's having less error than MSE+ADV. In terms of these dosimetric criteria, including the DVH loss has the best performance, even more than just including adversarial loss. Figure 8 is the prime example of how expert human domain knowledge can be used to greatly improve the model performance towards the domain relevant criteria. Since comformation, homogeneity, dose spillage, and dose coverage all use particular DVH values in its calculation, they all improved greatly from usage of $L_{DVH}$. Adversarial training for automatic learning of domain knowledge can further augment the performance by further capturing details that we did not specifically quantify in the loss. It does performs worse on its own, compared to the domain knowledge loss, due to the fact that the adversarial learning model does not truly know what is important to the user. However, when combined with our own domain knowledge, the performance can be maximized.

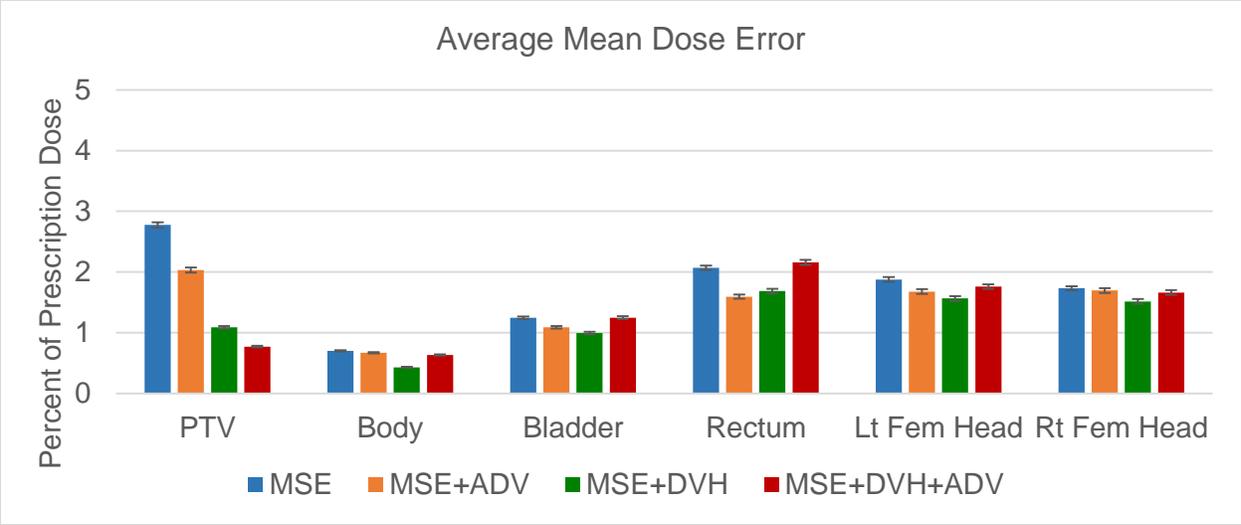

Figure 9: Average error in the mean dose for the PTV and the organs at risk. Error bar represents the 99% confidence interval $\left(\bar{x} \pm 2.576 * \frac{\sigma}{\sqrt{n}}\right)$, taken overall all test patients and plans.

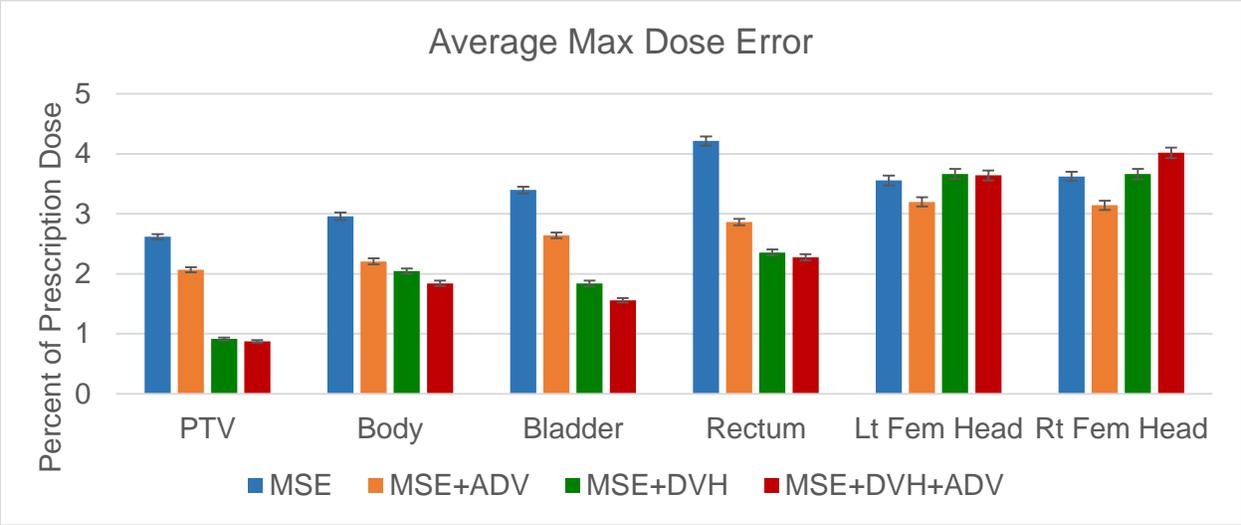

Figure 10: Average error in the max dose for the PTV and the organs at risk. Error bar represents the 99% confidence interval $\left(\bar{x} \pm 2.576 * \frac{\sigma}{\sqrt{n}}\right)$, taken overall all test patients and plans.

For both the mean dose PTV error and the max dose PTV, Body, Bladder and rectum error, the same improving trend can be observed in the order of the MSE model, MSE+ADV model, MSE+DVH model, and MSE+DVH+ADV model shown in Figures 9 and 10. However, there is not a clear trend in the mean dose for the OARs, due to the fact that MSE is already designed for reducing average errors, making it competitive for the mean dose error performance. There also lacks a trend for the max dose point for the femoral heads, which are further away from the PTV and are in the lower dose region that has higher variability in the dose distribution. All predictions have very low average mean and max dose errors of less than 2.8% and 4.2%, respectively.

Due to the large number of test plans we have found that, for conformity, homogeneity, and dose coverage (D95, D98, and D99), the MSE+DVH+ADV model had a statistically significant lower error than the other predictive models, with the largest p-value=0.007. For mean and max doses to the OARs, only 2 comparisons against the MSE+DVH+ADV model were found to be not significantly different, which was the mean dose error to the bladder against the MSE model (p-value=0.894), and the max dose error to the left femoral head against the MSE+DVH model (p-value=0.409). All other mean and max dose comparisons against the MSE+DVH+ADV had found statistically significant differences, with the largest p-value=0.042.

## IV. Discussion

To our knowledge, this is the first usage of a domain specific loss function, the DVH loss, for volumetric dose distribution prediction in radiation therapy. We compare the performance of deep neural networks trained using various loss combinations, including MSE loss, MSE+ADV loss, MSE+DVH loss, and MSE+DVH+ADV. Inclusion of the DVH loss had improved the model's prediction in almost every aspect, except for mean dose to the OARs and the max dose the femurs. The DVH loss does not directly represent mean or max dose, and thus is not directly minimizing these aspects. In addition, MSE loss is inherently designed to minimize average error, thus it is not surprising that MSE loss alone is competitive for driving the organ mean dose error down, since the additional DVH and ADV losses may have the model focus on aspects other than mean error. Regardless of the model, all predictions have an average mean and max dose error less than 2.8% and 4.2%, respectively, of the prescription dose for every structure of interest.

To be specific, the performance of our model improved with respect to our domain relevant metrics, because our domain knowledge-based losses are designed to reduce the error in very specific areas of the model's prediction, while focusing less on minimizing error for irrelevant metrics. In addition, having multiple losses can have a regularization effect, which can improve model generalizability and overall performance on unseen data. However, this does not guarantee that the model's performance will improve in all aspects, as indicated by the competitive organ mean dose error with the MSE model, since mean dose error is already heavily related to mean squared error.

Overall, the MSE+DVH+ADV performed the best in most of the categories, particularly the conformity, heterogeneity, high dose spillage, and planning target volume (PTV) dose coverage. This illustrates the importance of both human and learned domain knowledge. Expert human domain specific knowledge can greatly improve the performance of the model, tailoring the prediction towards domain relevant aspects. However, by having to explicit formulate this domain knowledge into an equation, it is difficult to capture the nuanced aspects of the problem. Using adversarial learning can then be used to further augment the model's performance, since the discriminator network can pick out the subtle aspects that the domain specific formulation may have missed.

Due to the non-convexity of both the DVH and ADV losses, as well as the inherent non-convex nature of neural networks, the MSE loss was utilized in every variant of the study, acting as the initial driving force and guide for the model to reasonably converge before the DVH and/or ADV losses began to take effect on the model's prediction. MSE loss still has many desirable properties from an optimization perspective. It is convex and has an extremely well behaved gradient. In addition the properties of the squared $\ell_2$-norm, where $\ell_p(x) = \sqrt[p]{\sum_i |x|^p}$, is one of the most understood and utilized functions in optimization[75]. It is not surprising that the previous studies achieved the state-of-the-art performance for dose prediction utilizing only MSE loss.

The final errors were assessed with 12,000 plans from 10 test patients, with varying tradeoff combinations. The total large number of plans with the randomization scheme given in Table 1 gives us confidence that the entire tradeoff space has been reasonably sampled. Theoretically, the space can be fully sampled from just using the "high weights" randomization scheme outlined in Table 1. However, it would take far more sampling points, since most of the plans deriving from this scheme would not be considered even close to clinically acceptable. By including weight randomizations in the "Single organ spare" category, we are able to create a particular single-organ-sparing plan, with varying degrees of sparing through randomization. Furthermore, the remaining randomization schemes allow for us to create general plans that are closer to clinical relevance than the "high weights" scheme. These effectively allow for us to smooth the tradeoff surface between the single-organ-sparing plans, and easily interpolate in between, making the total set of 12,000 plans representative of the different obtainable dose distribution.

The low prediction errors on the test patients signify that the model is capable of reliably generating Pareto optimal dose distributions with high accuracy. In addition, the raw prediction time of the neural network, including data movement to and from the GPU, is at 0.052 seconds. Realistically, with data loading, prediction, DVH calculation, and displaying the dose wash to a user, it takes approximately 0.6 seconds. This is still fast enough for real time interaction with the model to quickly explore the tradeoff space for a patient. The optimization based approach is much slower, first requiring, on average, 32 minutes for dose influence matrix calculation, and then 2 seconds for the optimization of each Pareto optimal dose. This allows for us to focus this tool towards empowering physicians. Immediately after segmentation of the cancer patient, the physician can immediately begin to generate a dose distribution with realistic and patient-specific tradeoffs between the PTV and various OARs. Not only does this give the physician a sense of the available and achievable tradeoffs, the resulting dose can then be given to a dosimetrist as a tangible and physician-preferred endpoint, alongside the other typical planning information provided by the physician. Usage of such a model is expected to drastically reduce the treatment planning time by reducing the number of feedback loops between the physician and dosimetrist, as well as how much the dosimetrist has to iterate through tuning hyperparameters in the fluence map optimization. The clinical relevance regarding the predictive improvement of the model—using MSE+DVH+ADV loses versus using only MSE—on the dosimetric constraints still require assessment through a clinical study to be properly quantified.

The addition of the adversarial loss increases the training time the most for training, since the discriminator has to be trained concurrently. The additional DVH loss does slow down the training as well, but has a much smaller effect than the adversarial loss. While the training times were wildly different, the final trained neural networks all yield the same exact prediction time, due to the fact that they have identical network architectures. The network that took the longest training time, MSE+DVH+ADV, took just under 4 days to train, which is still a very reasonable training time to prepare a model.

When training the multiple models, the weights $\lambda_{DVH}$ and $\lambda_{ADV}$, were chosen to be 0.1 and 0.001, respectively, in order to keep the losses at a similar order of magnitude until convergence. In general, this technique of assigning the human and learned domain knowledge weightings can be performed similarly. First the general loss model—for example, our MSE model—can be run first until convergence. The predictions of the general model can be used to assess its current loss value and the human domain knowledge metric to obtain orders of magnitude in the error. The adversarial model may have different orders of magnitude in its loss depending on the exact loss function used, but this may be estimated by bound evaluation of when the discriminator is able to perfectly distinguish the real data vs the predicted data and when it is unable to. The human and learned domain knowledge weightings can then be assigned by orders of magnitude, but some fine-tuning may be necessary depending on the problem to solve. From the clinical perspective, by setting the weights $\lambda_{DVH}$ and $\lambda_{ADV}$, such that the losses operate in a similar order of magnitude, we are setting equal importance to the overall general dose distribution (e.g. MSE), domain relevant metrics (e.g. DVH), and letting the network itself decide what is important (e.g. ADV). However, it may be necessary to fine tune the weightings, putting even more emphasis on a particular aspect for tackling a particular problem.

Since the invention and adoption of intensity modulated radiation therapy, optimization has become the backbone of radiation therapy systems. The user will place their ideal dose constraints in the system, and an inverse optimization process occurs to solve for the best treatment parameters to realize the dose distribution. While today's cost functions used commercially may be different than what is used in this study, the core concepts remain. We have our dose constraints, which is simplified in this study to have the PTV treated to prescription dose, and to minimize the dose to any organs-at risk. We also have structure importance weightings, $w_s$, where increasing this value for particular structure means to more heavily weight the imposed dose constraint. Typically, this means to further improve the dose coverage or homogeneity for the target, or to further reduce the dose for a particular critical structure. This study can be further extended in a future investigation where the optimization problem is replaced with a more complex formulation.

While this study was primarily focused on the evaluation of the DVH, ADV, and MSE losses, the final trained models do have their limitations. While these models are capable of generating dose distributions on the Pareto surface, it is currently limited to prostate cancer patients with 7 beam IMRT. In addition, the predicted dose distributions are not guaranteed to be deliverable, hence the current need for heavier dosimetrist involvement in the treatment planning. As a future study, we plan to broaden our Pareto optimal dose prediction work by improving our models to predict on different cancer sites, to handle a different number and orientation of beams for IMRT, and to work on the VMAT modality. In addition we plan to investigate the development of a fully automated treatment planning pipeline, starting with the implementation of a robust dose mimicking optimization engine, as the threshold-driven optimization for reference-based auto-planning (TORA) algorithm[25], which can be capable of generating a deliverable plan given a dose distribution or constraints. We expect such a pipeline would radically reduce the entire treatment planning time, especially for simple cases, allowing for the physician and dosimetrist to focus their efforts on more challenging patients.

# V. Conclusion

In this study, we proposed a novel domain specific loss function, the dose volume histogram (DVH) loss, and evaluated its efficacy alongside two other losses, mean squared error (MSE) loss and adversarial (ADV) loss. We trained and evaluated four instances of the models with varying loss combinations, which included 1) MSE, 2) MSE+ADV, 3) MSE+DVH, 4) MSE+DVH+ADV. We found that the models that included the domain specific DVH loss outperformed the models without the DVH loss in most of the categories, particularly on the evaluations of conformity, heterogeneity, high dose spillage, and planning target volume (PTV) dose coverage. The MSE+DVH+ADV model performed the best in these categories, illustrating the importance of both human and learned domain knowledge. Expert human domain specific knowledge can be the largest driver in the performance improvement, but it is difficult to capture nuanced aspects of the problem in an explicit formulation. Adversarial learning can be used to further capture these particular subtle attributes as part of the loss. The prediction of Pareto optimal doses can be performed in real-time, allowing for a physician to quickly navigate the tradeoff space for a patient, and produce a dose distribution as a tangible endpoint for the dosimetrist to use for planning. Eventually we plan to develop a fully automated treatment planning system. This is expected to considerably reduce the treatment planning time, while improving the treatment planning quality, allowing for clinicians to focus their efforts on the difficult and demanding cases.

# VI. Acknowledgements

This study was supported by the National Institutes of Health (NIH) R01CA237269 and the Cancer Prevention & Research Institute of Texas (CPRIT) IIRA RP150485.